\documentclass[journal,12pt,onecolumn]{IEEEtran}

\usepackage{booktabs} % For formal tables

\usepackage[english]{babel}
\usepackage{blindtext}
\usepackage{epsfig,endnotes,graphicx}
\usepackage{array}
\usepackage{subfigure}
\usepackage{url}
\usepackage{algorithm}
\usepackage{algpseudocode}
\usepackage{amsmath}
\usepackage{fancyhdr}

\usepackage{amssymb}

\DeclareMathOperator*{\argmin}{argmin}

\begin{document}

\title{Edge Intelligence: On-Demand Deep Learning Model Co-Inference with Device-Edge Synergy}

\author{En Li, Zhi Zhou, and Xu Chen \\School of Data and Computer Science \\ Sun Yat-sen University, Guangzhou, China
\thanks{En Li, Zhi Zhou, and Xu Chen, ``Edge Intelligence: On-Demand Deep Learning Model Co-Inference with Device-Edge Synergy,'' The SIGCOMM Workshop on Mobile Edge Communications, Budapest, Hungary, August 21-23, 2018.

The corresponding author is Xu Chen.

}}

\maketitle

\begin{abstract}
As the backbone technology of machine learning, deep neural networks (DNNs) have have quickly ascended to the spotlight. Running DNNs on resource-constrained mobile devices is, however, by no means trivial, since it incurs high performance and energy overhead. While offloading DNNs to the cloud for execution suffers unpredictable performance, due to the uncontrolled long wide-area network latency. To address these challenges, in this paper, we propose \textsf{Edgent}, a collaborative and on-demand DNN co-inference framework with device-edge synergy. \textsf{Edgent} pursues two design knobs: (1) DNN partitioning that adaptively partitions DNN computation between device and edge, in order to leverage hybrid computation resources in proximity for real-time DNN inference. (2) DNN right-sizing that accelerates DNN inference through early-exit at a proper intermediate DNN layer to further reduce the computation latency. The prototype implementation and extensive evaluations based on Raspberry Pi demonstrate \textsf{Edgent}'s effectiveness in enabling on-demand low-latency edge intelligence.
\end{abstract}

\section{Introduction \& Related Work}

As the backbone technology supporting modern intelligent mobile applications, Deep Neural Networks (DNNs)  represent the most commonly adopted machine learning technique and have become increasingly popular. Due to DNNs's ability to perform highly accurate and reliable inference tasks, they have witnessed successful applications in a broad spectrum of domains from computer vision \cite{SzegedyGoingCNN} to speech recognition \cite{Oord2016WaveNet} and natural language processing \cite{Wang2015LSTM}. However, as DNN-based applications typically require tremendous amount of computation, they cannot be well supported  by today's mobile devices with reasonable latency and energy consumption.

In response to the excessive resource demand of DNNs, the traditional wisdom resorts to the powerful cloud datacenter for training and evaluating DNNs. Input data generated from mobile devices is sent to the cloud for processing, and then results are sent back to the mobile devices after the inference. However, with such a cloud-centric approach, large amounts of data (e.g., images and videos) are uploaded to the remote cloud via a long wide-area network data transmission, resulting in high end-to-end latency and energy consumption of the mobile devices. To alleviate the latency and energy bottlenecks of cloud-centric approach, a better solution is to exploiting the emerging edge computing paradigm. Specifically, by pushing the cloud capabilities from the network core to the network edges (e.g., base stations and WiFi access points) in close proximity to devices, edge computing enables low-latency and energy-efficient DNN inference.

While recognizing the benefits of edge-based DNN inference, our empirical study reveals that the performance of edge-based DNN inference is highly sensitive to the available bandwidth between the edge server and the mobile device. Specifically, as the bandwidth drops from 1Mbps to 50Kbps, the latency of edge-based DNN inference climbs from 0.123s to 2.317s and becomes on par with the latency of local processing on the device. Then, considering the vulnerable and volatile network bandwidth in realistic environments (e.g., due to user mobility and bandwidth contention among various Apps), a natural question is that can we further improve the performance (i.e., latency) of edge-based DNN execution, especially for some mission-critical applications such as VR/AR games and robotics \cite{qualcomm2017arvr}.

To answer the above question in the positive, in this paper we proposed \textsf{Edgent}, a deep learning model co-inference framework with device-edge synergy. Towards low-latency edge intelligence\footnote{As an initial investigation, in this paper we focus on the execution latency issue. We will also consider the energy efficiency issue in a  future study.}, \textsf{Edgent} pursues two design knobs. The first is DNN partitioning, which adaptively partitions DNN computation between mobile devices and the edge server based on the available bandwidth, and thus to take advantage of the processing power of the edge server while reducing data transfer delay. However, worth noting is that the latency after DNN partition is still restrained by the rest part running on the device side. Therefore, \textsf{Edgent} further combines DNN partition with DNN right-sizing which accelerates DNN inference through early-exit at an intermediate DNN layer. Needless to say, early-exit naturally gives rise to the latency-accuracy tradeoff (i.e., early-exit harms the accuracy of the inference). To address this challenge,  \textsf{Edgent} jointly optimizes the DNN partitioning and right-sizing in an on-demand manner. That is, for mission-critical applications that typically have a pre-defined deadline, \textsf{Edgent} maximizes the accuracy without violating the deadline. The prototype implementation and extensive evaluations based on Raspberry Pi demonstrate \textsf{Edgent}'s effectiveness in enabling on-demand low-latency edge intelligence.

While the topic of edge intelligence has began to garner much attention recently, our study is different from and complementary to existing pilot efforts. On one hand, for fast and low power DNN inference at the mobile device side, various approaches as exemplified by DNN compression and DNN architecture optimization has been proposed \cite{Han2015Deep,Kim2016Compression,Wu2015Quantized,LaneDXTK,SqueezeNet}. Different from these works, we take a scale-out approach to unleash the  benefits of collaborative edge intelligence between the edge and mobile devices, and thus to mitigate the performance and energy bottlenecks of the end devices. On the other hand, though the idea of DNN partition among cloud and end device is not new \cite{Kang2017Neurosurgeon}, realistic measurements show that the DNN partition is not enough to satisfy the stringent timeliness requirements of mission-critical applications. Therefore, we further apply the approach of DNN right-sizing to speed up DNN inference.

\section{Background \& Motivation}
\label{motivation}
In this section, we first give a primer on DNN, then analyse the inefficiency of edge- or device-based DNN execution, and finally illustrate the benefits of DNN partitioning and right-sizing with device-edge synergy towards low-latency edge intelligence.
%\vspace{-5pt}
\subsection{A Primer on DNN}
\begin{figure}[!h]
  \centering
  %\vspace{-5pt}
  \includegraphics[scale=0.3]{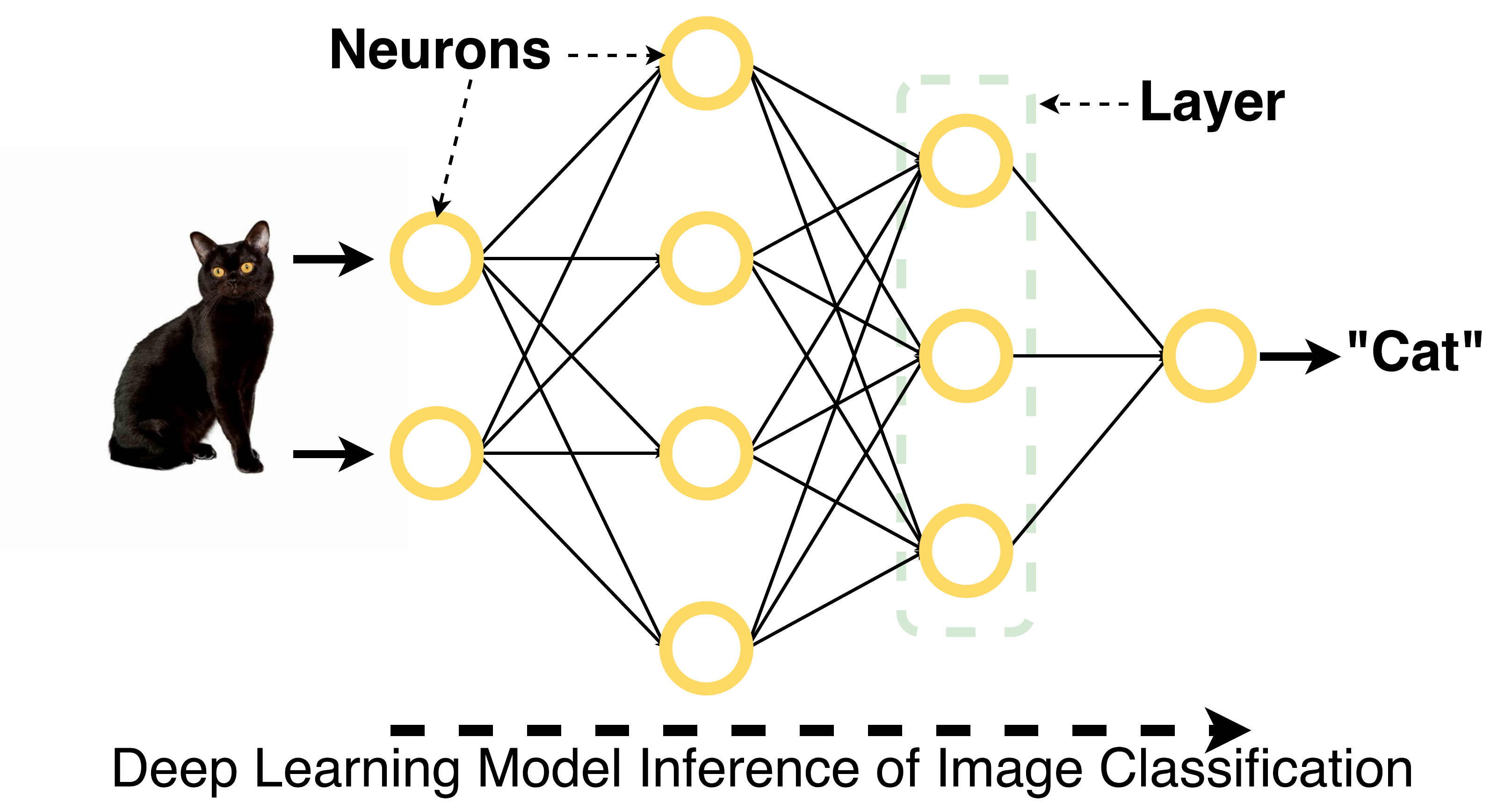}\\
  %\vspace{-10pt}
  \caption{A 4-layer DNN for computer vision}\label{dnnexample}
  %\vspace{-10pt}
\end{figure}
DNN represents the core machine learning technique for a broad spectrum of intelligent applications spanning computer vision, automatic speech recognition and natural language processing. As illustrated in Fig. \ref{dnnexample}, computer vision applications use DNNs to extract features from an input image and classify the image into one of the pre-defined categories. A typical DNN model is organized in a directed graph which includes a series of inner-connected layers, and within each layer comprising some number of nodes. Each node is a neuron that applies certain operations to its input and generates an output. The input layer of nodes is set by raw data while the output layer determines the category of the data. The process of passing forward from the input layer to the out layer is called model inference. For a typical DNN containing tens of layers and hundreds of nodes per layer, the number of parameters can easily reach the scale of millions. Thus, DNN inference is \emph{computational intensive}. Note that in this paper we focus on DNN inference, since DNN training is generally delay tolerant and is typically conducted in an off-line manner using powerful cloud resources.
%\vspace{-5pt}

%\vspace{-5pt}
\subsection{Inefficiency of Device- or Edge-based DNN Inference}

%\vspace{-8pt}
\begin{figure}[!h]
  \centering
  \includegraphics[scale=0.3]{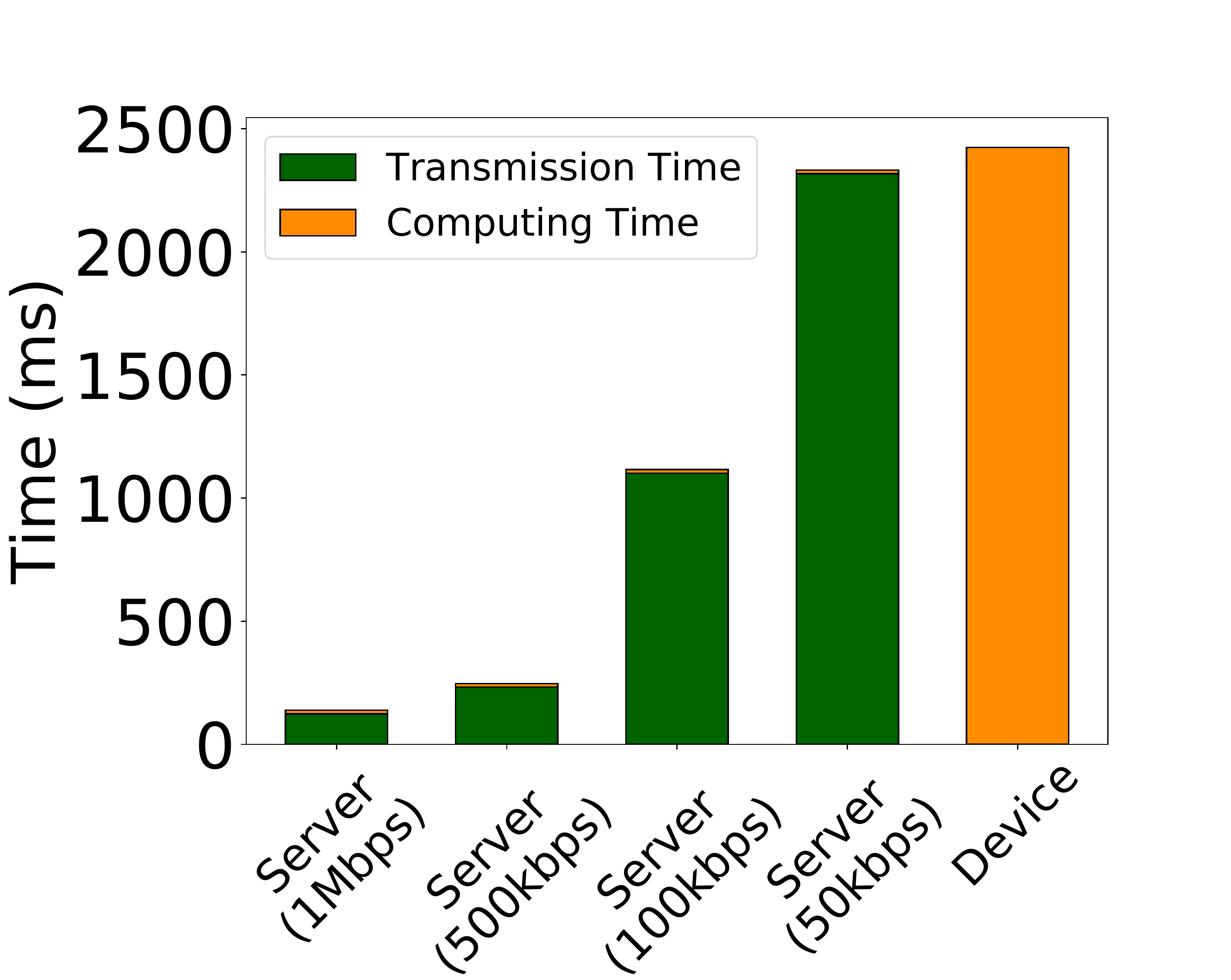}\\
  %\vspace{-10pt}
  \caption{AlexNet Runtime}\label{ineff}
  %\vspace{-13pt}
\end{figure}

Currently, the status quo of mobile DNN inference is either direct execution on the mobile devices or offloading to the cloud/edge server for execution. Unfortunately, both approaches may suffer from poor performance (i.e., end-to-end latency), being hard to well satisfy real-time intelligent mobile applications (e.g., AR/VR mobile gaming and intelligent robots) \cite{telecom2017mec}. As illustration, we take a Raspberry Pi tiny computer and a desktop PC to emulate the mobile device and edge server respectively, running the classical AlexNet \cite{NIPS2012_4824} DNN model for image recognition over the Cifar-10 dataset \cite{Krizhevsky2009Learning}. Fig. \ref{ineff} plots the breakdown of the end-to-end latency of different approaches under varying bandwidth between the edge and mobile device. It clearly shows that it takes more than 2s to execute the model on the resource-limited Raspberry Pi. Moreover, the performance of edge-based execution approach is dominated by the input data transmission time (the edge server computation time keeps at $\sim$10ms) and thus highly \emph{sensitive} to the available bandwidth. Specifically, as the available bandwidth jumps from 1Mbps to 50Kbps, the end-to-end latency climbs from 0.123s to 2.317s. Considering the network bandwidth resource scarcity in practice (e.g., due to network resource contention among users and apps) and computing resource limitation on mobile devices, both of the device- and edge-based approaches are challenging to well support many emerging real-time intelligent mobile applications with stringent latency requirement.

\subsection{Enabling Edge Intelligence with DNN Partitioning and Right-Sizing}

\begin{figure}[!h]
  \centering
  \includegraphics[scale=0.3]{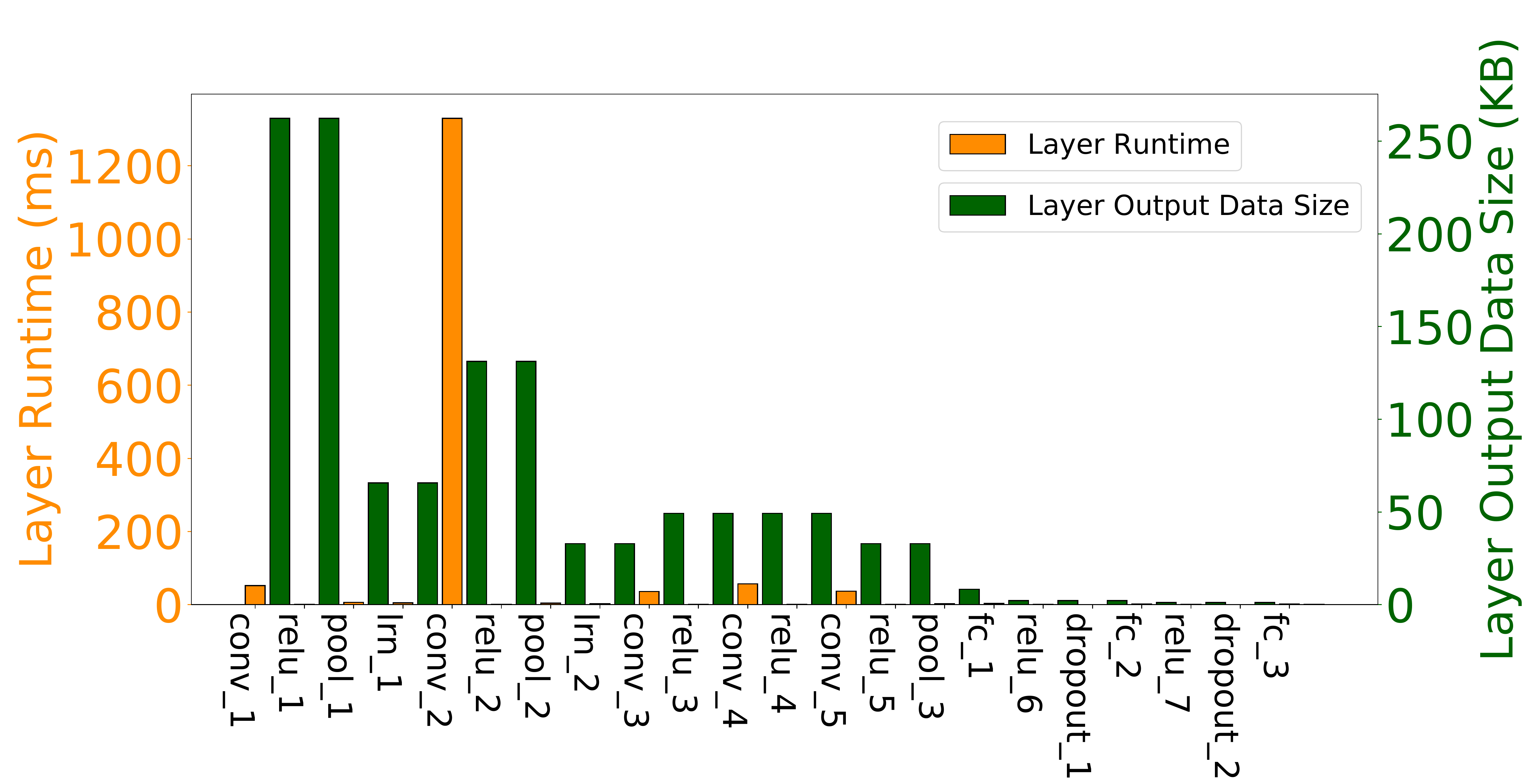}\\
  %\vspace{-10pt}
  \caption{AlexNet Layer Runtime on Raspberry Pi}\label{Runtime}
  %\vspace{-15pt}
\end{figure}

\textbf{DNN Partitioning:} For a better understanding of the performance bottleneck of DNN execution, we further break the runtime (on Raspberry Pi) and output data size of each layer in Fig. \ref{Runtime}. Interestingly, we can see that the runtime and output data size of different layers exhibit great heterogeneities, and layers with a long runtime do not necessarily have a large output data size. Then, an intuitive idea is DNN partitioning, i.e., partitioning the DNN into two parts and offloading the computational intensive one to the server at a low transmission overhead, and thus to reduce the end-to-end latency. For illustration, we choose the second local response normalization layer (i.e., \texttt{lrn\_2}) in Fig. \ref{Runtime} as the partition point and offload the layers before the partition point to the edge server while running the rest layers on the device. By DNN partitioning between device and edge, we are able to collaborate hybrid computation resources in proximity for low-latency DNN inference.%Experiments show that as the available bandwidth varies from 1Mbps to 50Kbps, the latency with DNN partition is reduced to 0.6$\sim$4.59s, meaning a speedup of -1.89$\sim$4.04$X$ compared to the edge-based approach.

\begin{figure}[!h]
	\centering
	\includegraphics[scale=0.3]{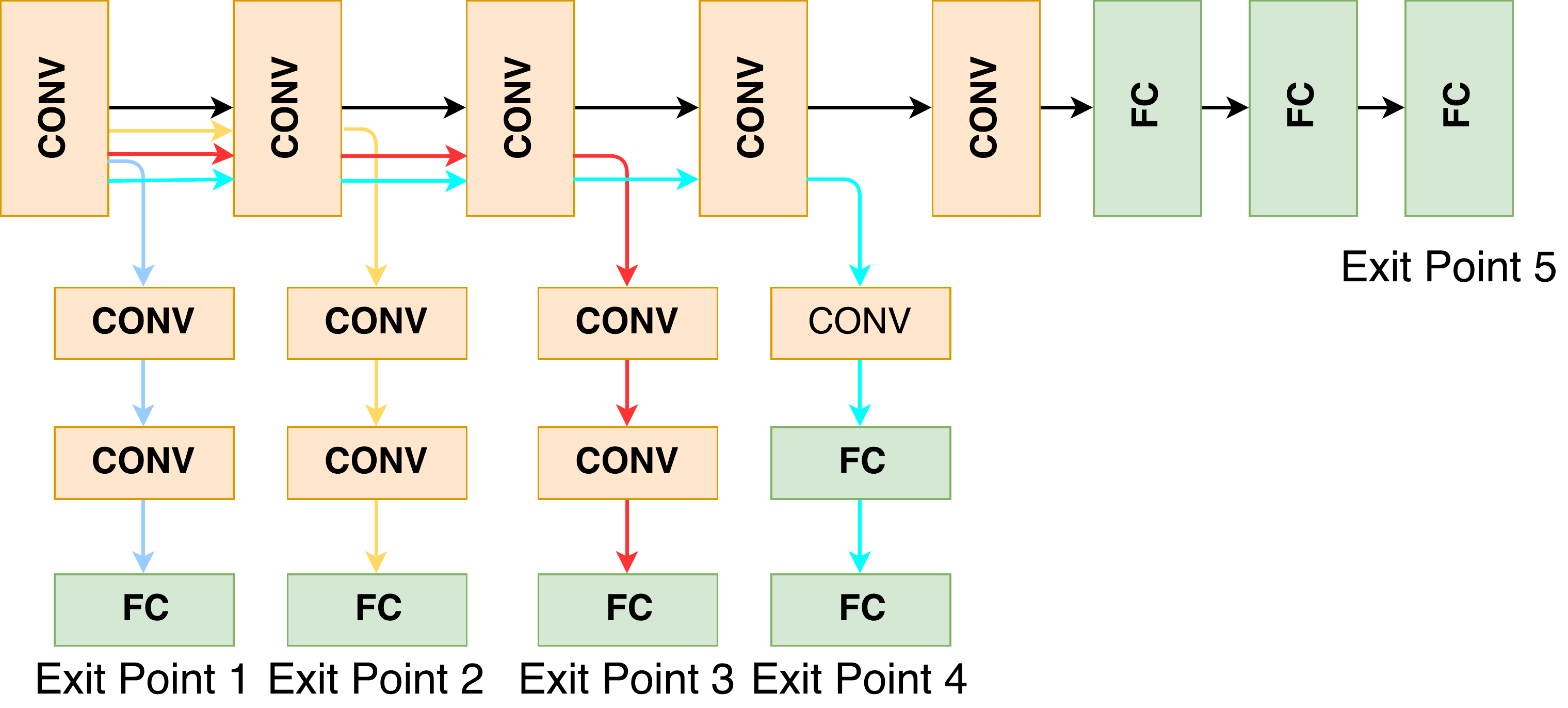}\\
	%\vspace{-10pt}
	\caption{An illustration of the early-exit mechanism for DNN right-sizing}\label{Branchy AlexNet}
	%\vspace{-10pt}
\end{figure}

\textbf{DNN Right-Sizing:} While DNN partitioning greatly reduces the latency by bending the computing power of the edge server and mobile device, we should note that the optimal DNN partitioning is still constrained by the run time of layers running on the device. For further reduction of latency, the approach of \emph{DNN Right-Sizing} can be combined with DNN partitioning. DNN right-sizing promises to accelerate model inference through an early-exit mechanism. That is, by training a DNN model with multiple exit points and each has a different size, we can choose a DNN with a small size tailored to the application demand, meanwhile to alleviate the computing burden at the model division, and thus to reduce the total latency. Fig. \ref{Branchy AlexNet} illustrates a branchy AlexNet with five exit points. Currently, the early-exit mechanism has been supported by the open source framework BranchyNet\cite{7900006}. Intuitively, DNN right-sizing further reduces the amount of computation required by the DNN inference tasks.%Our experiments show that by combining DNN right-sizing to DNN partitioning, the latency can be further reduced by 1.06$X$ compared to the partitioning-only approach (given a bandwidth of 500Kbps).

\textbf{Problem Description:} Obviously, DNN right-sizing incurs the problem of latency-accuracy tradeoff --- while early-exit reduces the computing time and the device side, it also deteriorates the accuracy of the DNN inference. Considering the fact that some applications (e.g., VR/AR game) have stringent deadline requirement while can tolerate moderate accuracy loss, we hence strike a nice balance between the latency and the accuracy in an on-demand manner. Particularly,  given the predefined and stringent latency goal, we maximize the accuracy without violating the deadline requirement. More specifically, the problem to be addressed in this paper can be summarized as: given a predefined latency requirement, how to jointly optimize the decisions of DNN partitioning and right-sizing, in order to maximize DNN inference accuracy.
%\vspace{-8pt}

%\vspace{-5pt}
\section{Framework}

We now outline the initial design of \textsf{Edgent}, a framework that automatically and  intelligently selects the best partition point and exit point of a DNN model to maximize the accuracy while satisfying the requirement on the execution latency.

%\vspace{-5pt}
\subsection{System Overview}

\begin{figure*}[!ht]
  \centering
  \includegraphics[scale=0.7]{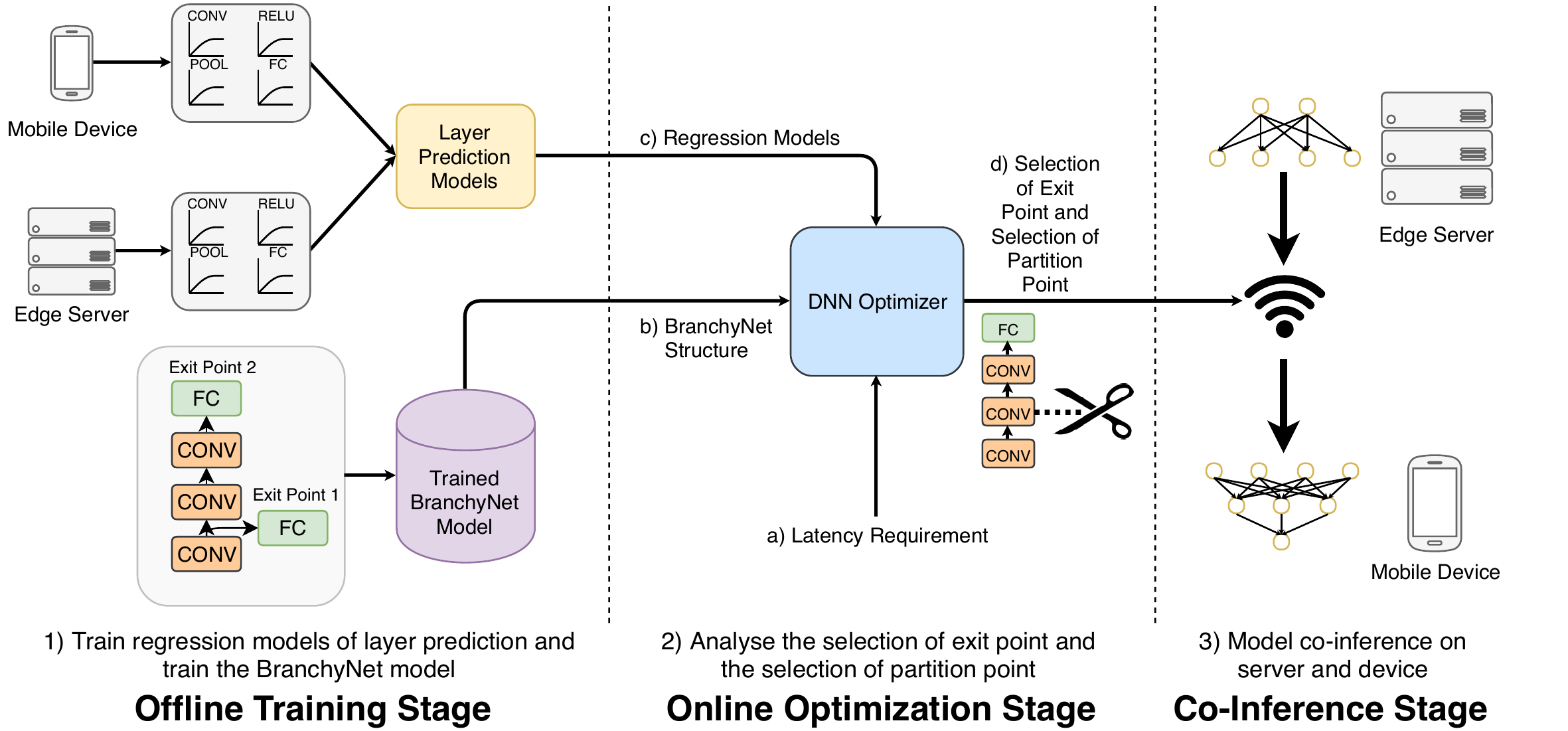}\\
  %\vspace{-10pt}
  \caption{Edgent overview}\label{Edgent Overview}
  %\vspace{-5pt}
\end{figure*}

Fig. \ref{Edgent Overview} shows the overview of \textsf{Edgent}. \textsf{Edgent} consists of three stages: offline training stage, online optimization stage and co-inference stage.

At \textbf{offline training stage}, \textsf{Edgent} performs two initializations: (1) profiling the mobile device and the edge server to generate regression-based performance prediction models (Sec. \ref{performance-predict}) for different types DNN layer (e.g., \texttt{Convolution}, \texttt{Pooling}, etc.). (2) Using \texttt{Branchynet} to train DNN models with various exit points, and thus to enable early-exit. Note that the performance profiling is infrastructure-dependent, while the DNN training is application-dependent. Thus, given the sets of infrastructures (i.e., mobile devices and edge servers) and applications, the two initializations only need to be done once in an offline manner.

At \textbf{online optimization stage}, the DNN optimizer selects the best partition point and early-exit point of DNNs to maximize the accuracy while providing performance guarantee on the end-to-end latency, based on the input: (1) the profiled layer latency prediction models and Branchynet-trained DNN models with various sizes. (2) the observed available bandwidth between the mobile device and edge server. (3) The pre-defined latency requirement. The optimization algorithm is detailed in Sec. \ref{opt}.

At \textbf{co-inference stage}, according to the partition and early-exit plan, the edge server will execute the layer before the partition point and the rest will run on the mobile device.
\vspace{-10pt}
\subsection{Layer Latency Prediction} \label{performance-predict}

\begin{table}[]
\centering
\caption{The independent variables of regression models}
%\vspace{-10pt}
\label{independent var}
\footnotesize
%\small
\begin{tabular}{>{\centering\arraybackslash}m{2.3cm}|>{\centering\arraybackslash}m{4.9cm}}
\hline
\textbf{Layer Type} & \textbf{Independent Variable} \\ \hline
Convolution & \begin{tabular}[c]{@{}l@{}}amount of input feature maps,\\ (filter size/stride)\textasciicircum 2*(num of filters)\end{tabular}\\ \hline
Relu & input data size \\ \hline
Pooling & \begin{tabular}[c]{@{}l@{}}input data size, output data size\end{tabular} \\ \hline
Local Response Normalization & input data size \\ \hline
Dropout & input data size \\ \hline
Fully-Connected & \begin{tabular}[c]{@{}l@{}}input data size, output data size\end{tabular} \\ \hline
Model Loading & model size \\ \hline
\end{tabular}
\end{table}

When estimating the runtime of a DNN, \textsf{Edgent} models the per-layer latency rather than modeling at the granularity of a whole DNN. This greatly reduces the profiling overhead since there are very limited classes of layers. By experiments, we observe that the latency of different layers is determined by various independent variables (e.g., input data size, output data size) which are summarized in Table \ref{independent var}. Note that we also observe that the DNN model loading time also has an obvious impact on the overall runtime. Therefore, we further take the DNN model size as a input parameter to predict the model loading time. Based on the above inputs of each layer, we establish a regression model to predict the latency of each layer based on its profiles. The final regression models of some typical layers are shown in Table \ref{rg model} (size is in bytes and latency is in ms).

%\vspace{-5pt}
%\vspace{45pt}

\begin{table*}[!ht]
\centering
\caption{Regression model of each type layer}
%\vspace{-10pt}
\label{rg model}
\footnotesize
\begin{tabular}{>{\centering\arraybackslash}m{2.3cm}|c|c}
\hline
\textbf{Layer}   & \textbf{Mobile Device model}                               & \textbf{Edge Server model}                             \\ \hline
Convolution    & y = 6.03e-5 $\ast$ x1 + 1.24e-4 $\ast$ x2 + 1.89e-1   & y = 6.13e-3 $\ast$ x1 + 2.67e-2 $\ast$ x2 - 9.909 \\
Relu    & y = 5.6e-6 $\ast$ x + 5.69e-2                    & y = 1.5e-5 $\ast$ x + 4.88e-1                  \\
Pooling    & y = 1.63e-5 $\ast$ x1 + 4.07e-6 $\ast$ x2 + 2.11e-1 & y = 1.33e-4 $\ast$ x1 + 3.31e-5 $\ast$ x2 + 1.657 \\
Local Response Normalization     & y = 6.59e-5 $\ast$ x + 7.80e-2                    & y = 5.19e-4 $\ast$ x+ 5.89e-1                    \\
Dropout & y = 5.23e-6 $\ast$ x+ 4.64e-3                    & y = 2.34e-6 $\ast$ x+ 0.0525                    \\
Fully-Connected      & y = 1.07e-4 $\ast$ x1 - 1.83e-4 $\ast$ x2 + 0.164 & y = 9.18e-4 $\ast$ x1 + 3.99e-3 $\ast$ x2 + 1.169  \\
Model Loading      & y = 1.33e-6 $\ast$ x + 2.182                  & y = 4.49e-6 $\ast$ x + 842.136 \\ \hline
\end{tabular}
\end{table*}
%\vspace{-1pt}

\subsection{Joint Optimization on DNN Partition and DNN Right-Sizing} \label{opt}
At online optimization stage, the DNN optimizer receives the latency requirement from the mobile device, and then searches for the optimal exit point and partition point of the trained branchynet model. The whole process is given in Algorithm 1. For a branchy model with $M$ exit points,  we denote that the $i\text{-}th$ exit point has $N_i$ layers. Here a larger index $i$ correspond to a more accurate inference model of larger size. We use the above-mentioned regression models to predict $ED_{j}$ the runtime of the $j\text{-}th$ layer when it runs on device and $ES_{j}$ the runtime of the $j\text{-}th$ layer when it runs on server. $D_{p}$ is the output of the $p\text{-}th$ layer. Under a specific bandwidth $B$, with the input data $Input$, then we calculate $A_{i, p}$ the whole runtime $(\sum_{j=1}^{p-1}ES_{j} + \sum_{k=p}^{N_{i}}ED_{j} + Input/B + D_{p-1}/B)$ when the $p\text{-}th$ is the partition point of the selected model of $i\text{-}th$ exit point.When $p=1$, the model will only run on the device then $ES_{p}=0, D_{p-1}/B=0, Input/B=0$ and when $p=N_{i}$, the model will only run on the server then $ED_{p}=0, D_{p-1}/B=0$. In this way, we can find out the best partition point having the smallest latency for the model of $i\text{-}th$ exit point. Since the model partition does not affect the inference accuracy, we can then sequentially try the DNN inference models with different exit points(i.e., with different accuracy), and find the one having the largest size and meanwhile satisfying the latency requirement. Note that since regression models for layer latency prediction are trained beforehand, Algorithm 1 mainly involves linear search operations and can be done very fast (no more than 1ms in our experiments).

\begin{algorithm}[htp]
%\scriptsize
\small
\caption{Exit Point and Partition Point Search}
\begin{flushleft}
\hspace*{0.02in} {\bf Input:}
\\ $M$: number of exit points in a branchy model
\\ $\{N_{i}|i=1,\cdots ,M \}$: number of layers in each exit point
\\ $\{L_{j}|j=1,\cdots ,N_{i}\}$: layers in the $i\text{-}th$ exit point
\\ $\{D_{j}|j=1,\cdots ,N_{i}\}$: each layer output data size of $i\text{-}th$ exit point
\\ $f(L_{j})$: regression models of layer runtime prediction
\\ $B$: current wireless network uplink bandwidth
\\ $Input$: the input data of the model
\\ $latency$: the target latency of user requirement\\
\hspace*{0.02in} {\bf Output:}
\\ Selection of exit point and its partition point
\end{flushleft}

\begin{algorithmic}[1]
\State \textbf{Procedure}
\For{$i=M,\cdots ,1$}
    \State Select the $i\text{-}th$ exit point
    \For{$j = 1,\cdots ,N_{i}$}
    \State $ES_{j}\gets f_{edge}(L_{j})$
    \State $ED_{j}\gets f_{device}({L_{j}})$
    \EndFor
    \State $A_{i, p} = \argmin\limits_{p=1,\cdots ,N_{i}} (\sum_{j=1}^{p-1}ES_{j} + \sum_{k=p}^{N_{i}}ED_{j} + Input/B + D_{p-1}/B)$
%\EndFor
%\For{$i=M,\cdots ,1$}
    \If{$A_{i, p} \leq latency$}
        \State \Return Selection of Exit Point $i$ and its Partition Point $p$
    \EndIf
\EndFor
\State \Return NULL \Comment can not meet latency requirement
\end{algorithmic}
\end{algorithm}

\begin{figure}[!ht]
  \centering
  \includegraphics[scale=0.4]{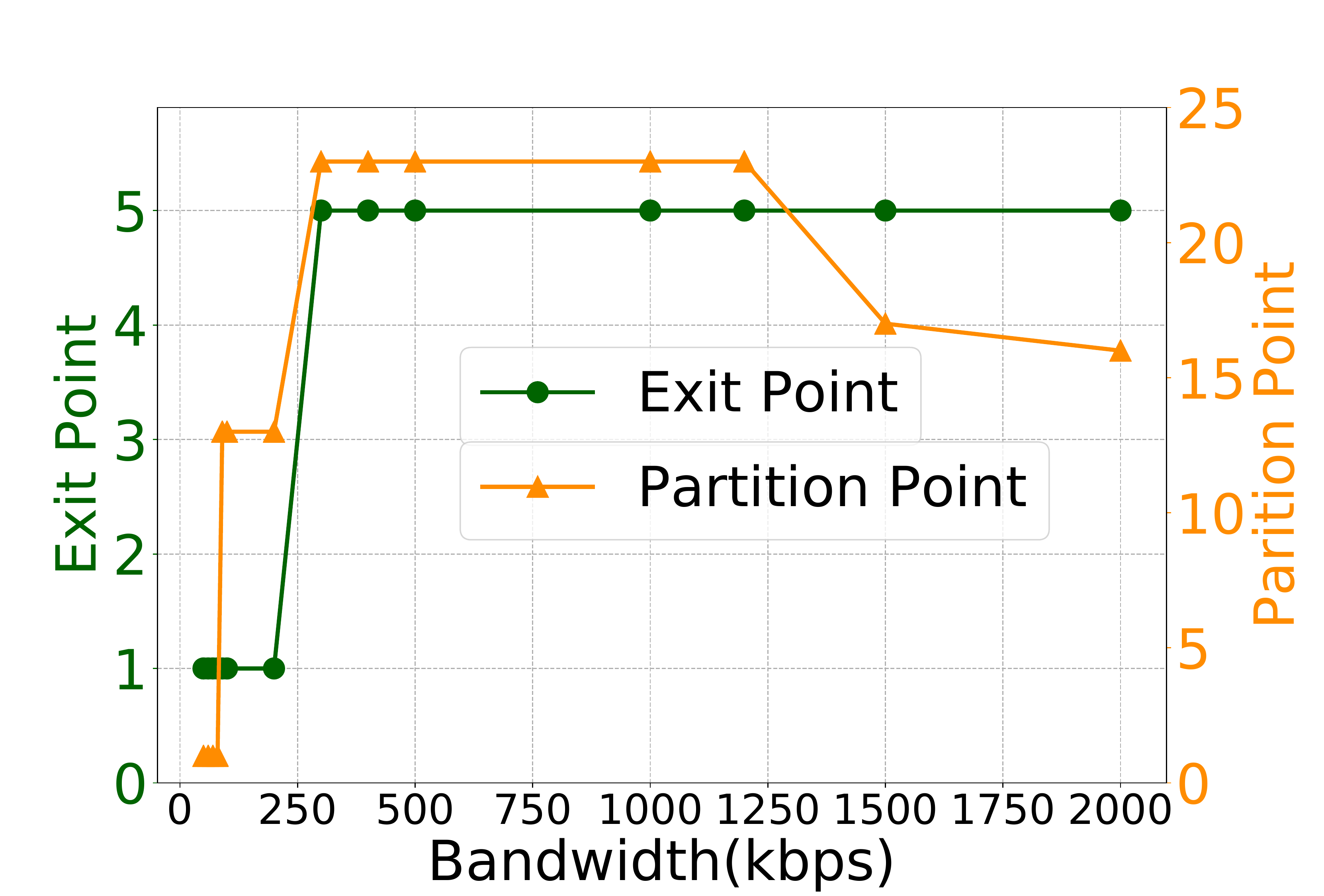}
  %\vspace{-10pt}
  \caption{Selection under different bandwidths}\label{exit and partition}
\end{figure}

%\begin{figure}[!ht]
%  \centering
%  % Requires \usepackage{graphicx}
%  \includegraphics[scale=0.158]{low_bandwidth_latency.pdf}
%  \caption{Model runtime under different bandwidths}\label{exit and partition}
%\end{figure}

\begin{figure}[!ht]
  \centering
  \includegraphics[scale=0.4]{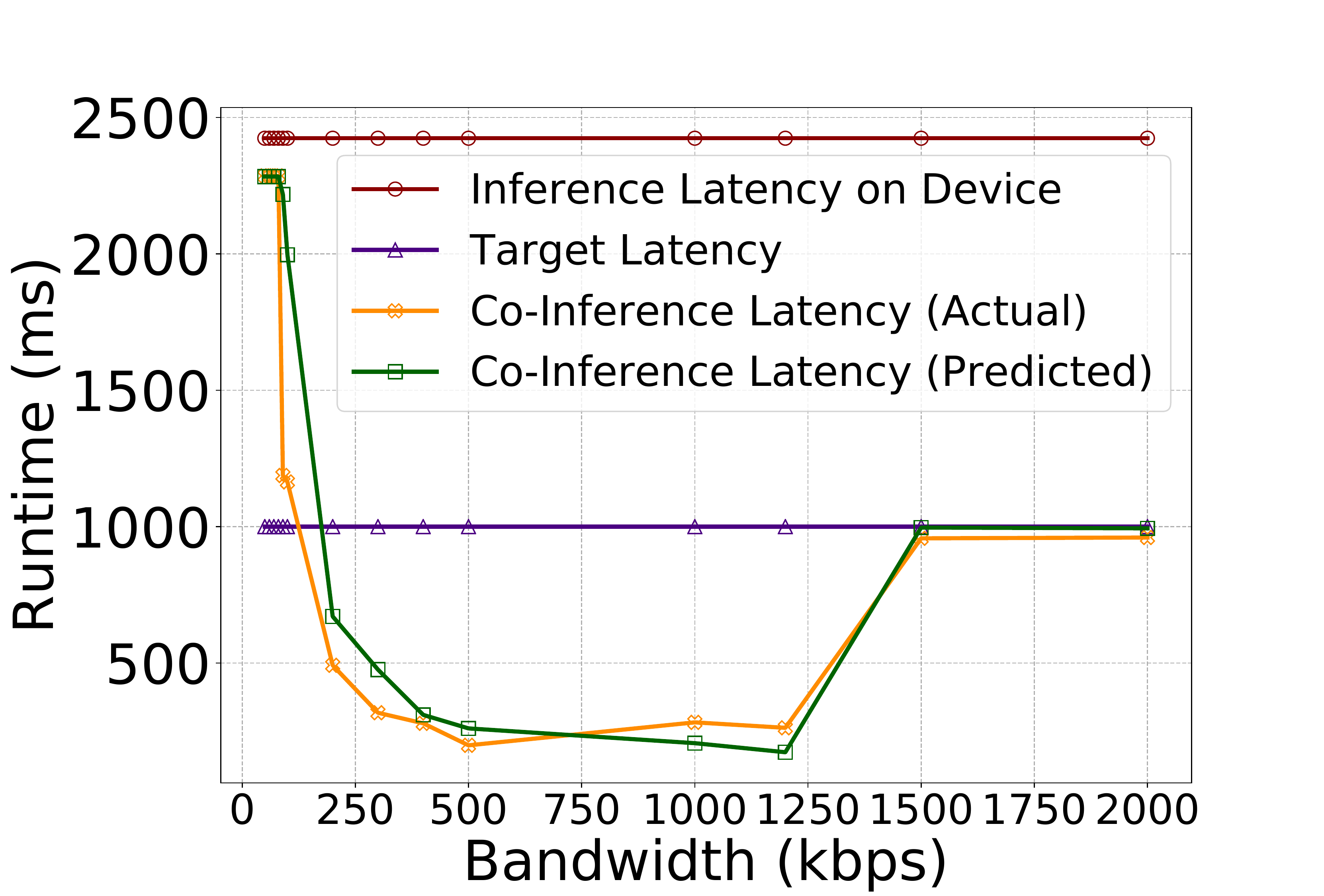}
  \caption{Model runtime under different bandwidths}\label{model runtime}
\end{figure}

\begin{figure}[!ht]
  \centering
  \includegraphics[scale=0.4]{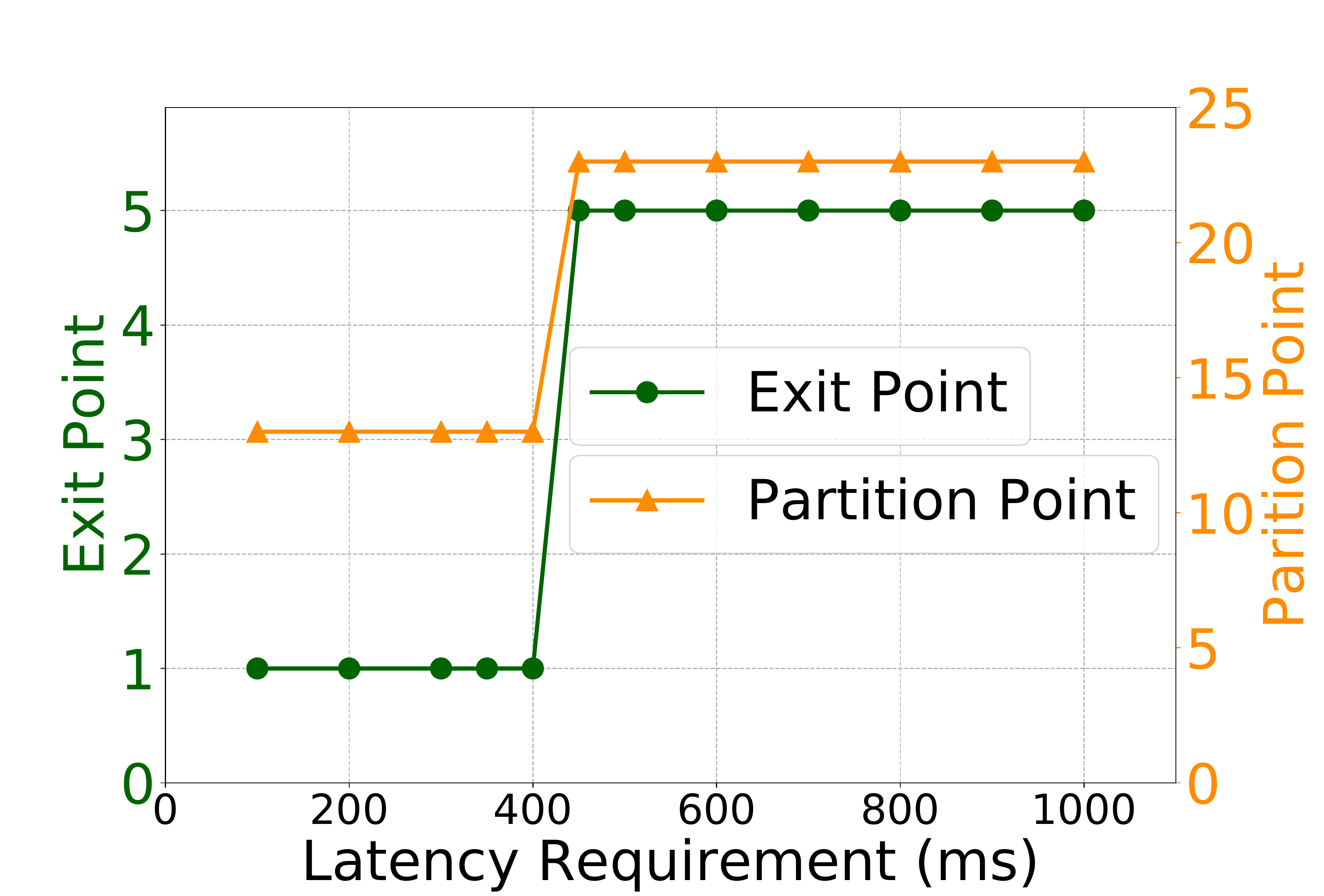}
  \caption{Selection under different latency requirements}\label{exit and partition diff latency}
\end{figure}

%\begin{figure*}[!ht]
%  \centering
%  % Requires \usepackage{graphicx}
%
%  \subfigure[Selection under different bandwidths]{
%  \includegraphics[scale=0.158]{low_bandwidth.pdf}\label{exit and partition}
%  }
%  \subfigure[Model runtime under different bandwidths]{
%  \includegraphics[scale=0.158]{low_bandwidth_latency.pdf}\label{model runtime}
%  }
%  \subfigure[Selection under different latency requirements]{
%  \includegraphics[scale=0.158]{500kbps_diff_latency.pdf}\label{exit and partition diff latency}
%  }
%  %\vspace{-10pt}
%  \caption{The results under different bandwidths and different latency requirements}\label{selections}
%  %\vspace{-10pt}
%\end{figure*}

\section{Evaluation}

We now present our preliminary implementation and evaluation results.

%\vspace{-1pt}
\subsection{Prototype}

\par We have implemented a simple prototype of \textsf{Edgent} to verify the  feasibility and efficacy of our idea. To this end, we take a desktop PC to emulate the edge server, which is equipped with a quad-core Intel processor at 3.4 GHz with 8 GB of RAM, and runs the Ubuntu system. We further use Raspberry Pi 3 tiny computer to act as a mobile device. The Raspberry Pi 3 has a quad-core ARM processor at 1.2 GHz with 1 GB of RAM. The available bandwidth between the edge server and the mobile device is controlled by the WonderShaper \cite{wondershaper} tool. As for the deep learning framework, we choose Chainer \cite{chainer} that can well support branchy DNN structures.

\par For the branchynet model, based on the standard AlexNet model, we train a branchy AlexNet for image recognition over the large-scale Cifar-10 dataset \cite{Krizhevsky2009Learning}. The branchy AlexNet has five exit points as showed in Fig. \ref{Branchy AlexNet} (Sec. \ref{motivation}), each exit point corresponds to a sub-model of the branchy AlexNet. Note that in Fig. \ref{Branchy AlexNet}, we only draw the convolution layers and the fully-connected layers but ignore other layers for ease of illustration. For the five sub-models, the number of layers they each have is 12, 16, 19, 20 and 22, respectively.

\par For the regression-based latency prediction models for each layer, the independent variables are shown in the Table. \ref{independent var}, and the obtained regression models are shown in Table \ref{rg model}.

\subsection{Results}

\par We deploy the branchynet model on the edge server and the mobile device to evaluate the performance of \textsf{Edgent}. Specifically, since both the pre-defined latency requirement and the available bandwidth play vital roles in \textsf{Edgent}'s optimization logic, we evaluate the performance of \textsf{Edgent} under various latency requirements and available bandwidth.

\par We first investigate the effect of the bandwidth by fixing the latency requirement at 1000ms and varying the bandwidth 50kbps to 1.5Mbps. Fig. \ref{exit and partition} shows the best partition point and exit point under different bandwidth. While the best partition points may fluctuate, we can see that the best exit point gets higher as the bandwidth increases. Meaning that the higher bandwidth leads to higher accuracy. Fig. \ref{model runtime} shows that as the bandwidth increases, the model runtime first drops substantially and then ascends suddenly. However, this is reasonable since the accuracy gets better while the latency is still within the latency requirement when increase the bandwidth from 1.2Mbps to 2Mbps. It  also shows that our proposed regression-based latency approach can well estimate the actual DNN model runtime latency. We further fix the bandwidth at 500kbps and vary the latency from 100ms to 1000ms. Fig. \ref{exit and partition diff latency} shows the best partition point and exit point under different latency requirements. As expected, the best exit point gets higher as the latency requirement increases, meaning that a larger latency goal gives more room for accuracy improvement.

\begin{figure}[!ht]
  \centering
  \includegraphics[scale=0.3]{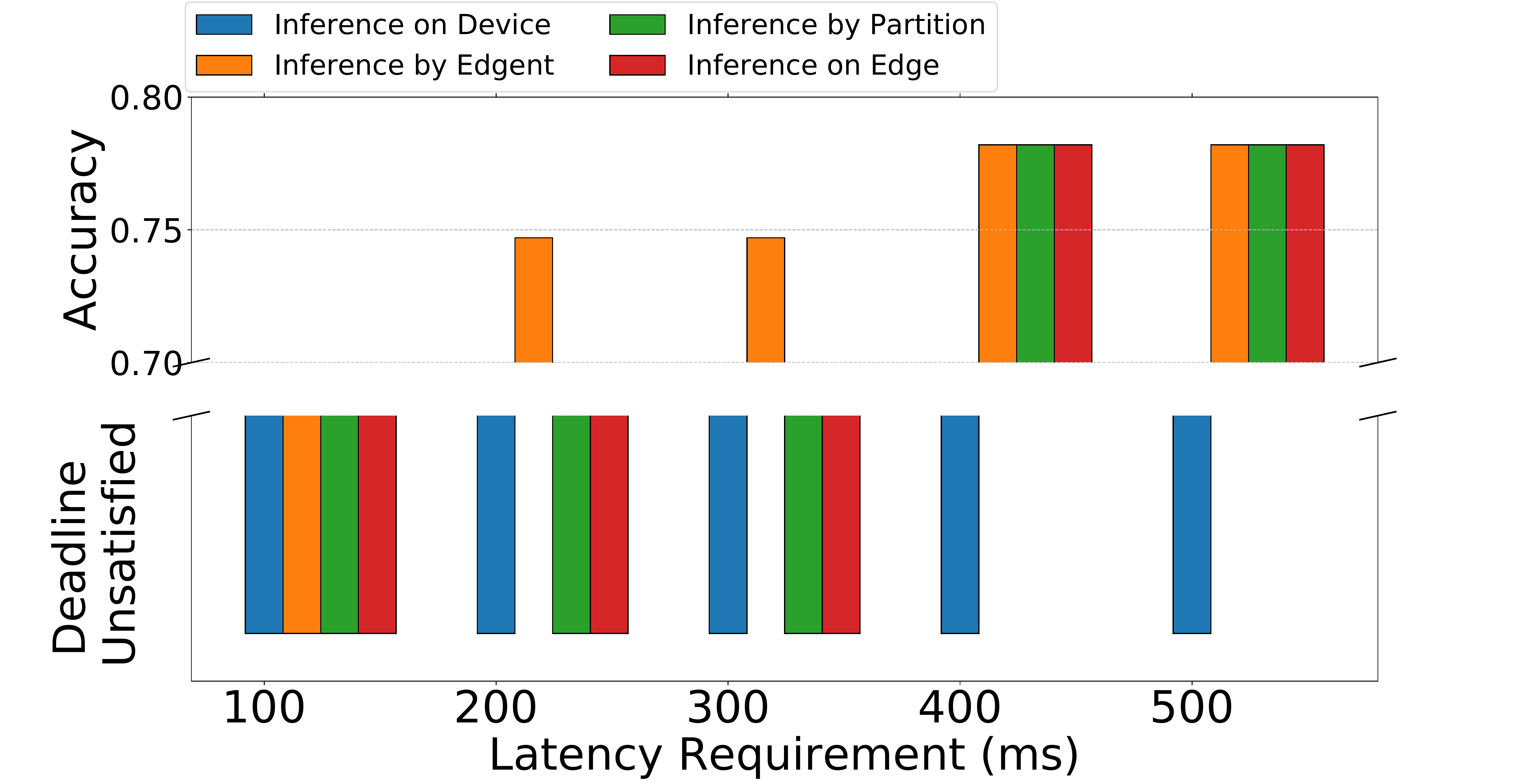}\\
  %\vspace{-10pt}
  \caption{The accuracy comparison under different latency requirement}\label{accuracy comparison}
\end{figure}

\par In Fig. \ref{accuracy comparison}, under different latency requirements, it shows the model accuracy of different inference methods. The accuracy is negative if the inference can not satisfy the latency requirement. The network bandwidth is set to 400kbps. Seen in the Fig. \ref{accuracy comparison}, at a very low latency requirement (100ms), all four methods can't satisfy the requirement. As the latency requirement increases, inference by \textsf{Edgent} works earlier than the other methods that at the 200ms to 300ms requirements, by using a small model with a moderate inference accuracy to meet the requirements. The accuracy of the model selected by \textsf{Edgent} gets higher as the latency requirement relaxes.

\section{Conclusion}

In this work, we present \textsf{Edgent}, a collaborative and on-demand DNN co-inference framework with device-edge synergy. Towards low-latency edge intelligence, \textsf{Edgent} introduces two design knobs to tune the latency of a DNN model: DNN partitioning which enables the collaboration between edge and mobile device, and DNN right-sizing which shapes the computation requirement of a DNN. Preliminary implementation and evaluations on Raspberry Pi demonstrate the effectiveness of \textsf{Edgent} in enabling low-latency edge intelligence. Going forward, we hope more discussion and effort can be
stimulated in the community to fully accomplish the vision of intelligent edge service.

\bibliographystyle{ieeetran}
%\bibliography{sample-bibliography}
\bibliography{reference}
\end{document}